\documentclass[
    ,final            
  ]
  {aipproc}
\usepackage{amsmath}
\usepackage{amssymb}
\layoutstyle{6x9}

\makeatother

\def\ket#1{\left|#1\right>}

\begin{document}

\title{SU(2) and SU(4) Kondo effect in double quantum dots}


\author{Jernej Mravlje}{
  address={Jo\v{z}ef Stefan Institute, Jamova 39, Ljubljana, Slovenia}
}

\author{Anton Ram\v{s}ak}{
  address={Faculty of Mathematics and Physics, University of
  Ljubljana, Jadranska 19, Ljubljana, Slovenia}
  ,altaddress={Jo\v{z}ef Stefan Institute, Jamova 39, Ljubljana, Slovenia} 
}

\author{Toma\v{z} Rejec}{
  address={Faculty of Mathematics and Physics, University of
  Ljubljana, Jadranska 19, Ljubljana, Slovenia}
  ,altaddress={Jo\v{z}ef Stefan Institute, Jamova 39, Ljubljana, Slovenia} 
}

\begin{abstract}
  We investigate serial double quantum dot systems with on-site and
  inter-site interaction by means of Sch\"onhammer-Gunnarsson
  projection-operator method. The ground state is established by the 
  competition between extended Kondo phases and localized singlet phases in spin and charge
 degrees of freedom. We present and discuss different
 phases, as discerned by characteristic correlation functions. We
 discuss also how different phases would be seen in linear transport
 measurements.
 
\end{abstract}

\maketitle


\section{INTRODUCTION}
\indent

In the last decade the advances in experimental techniques enabled the
exploration of intriguing many-body effects occurring in solid-state
systems such as the Kondo effect \cite{hewson_book} by means of
measuring the conductance of nanoscale electrical circuits. Tiny
pools of electrons defined by electrodes -- quantum dots (QDs) --
constitute artificial atoms/molecules. Additional gates 
enable tuning of the orbital levels as well as the tunneling rates,
which makes systematic exploration of  
various effects experimentally accessible. The Kondo effect is
essentially the increased scattering rate 
(with phase shifts near $\pi/2$) at low temperatures due to magnetic
impurities in host metals. In transport experiments through quantum
dots it is seen in another disguise: it is discerned as the
amplification of the conductance towards unitary 
limit. Interesting way to proceed further is to analyze the
consequences of inter-impurity interaction by looking at the
transport through double quantum dot (DQD) systems.

The characteristic feature of the two-impurity Kondo physics 
is that the two impurities
either form an inter-impurity singlet, which is virtually decoupled from
conduction electrons or they form a double Kondo state
SU(2)$\times$SU(2), in which each spin characterized by the SU(2) symmetry
group is screened by the conduction electrons \cite{jones88} depending
on the scales of the energies of the 
inter-impurity singlet formation $J$ and Kondo state formation
$T_K$. When the symmetry of the Hamiltonian is larger the Kondo
temperature is enhanced. For double quantum dots, which have the
capacitative interaction $V$ tuned near the value of the on-dot
interaction $U$, the SU(4) Kondo effect occurs \cite{galpin05}.   

Here we report our results on the competition between extended Kondo and
localized singlet phases in serial DQD systems with inter-dot
interaction in the point of particle-hole symmetry \cite{mravlje06}
and discuss also the 
phases which occur outside this point. The SU(4) Kondo
phase cannot be explored directly by transport experiment through a
DQD as the conductance is small irrespective of whether the system is
in the SU(4) Kondo state or not. Nevertheless, the scale of the SU(4) condensation energy
can be estimated  by tuning the system away from the point of SU(4)
symmetry until the SU(4) Kondo state collapses. The boundary
is easy to discern from the conductance data  as the conductance is
unity whenever the crossover between the phases takes place.

\section{MODEL  AND METHOD}
We model DQDs by the two-impurity Anderson
Hamiltonian $H=H_{\mathrm{d}}+H_{\mathrm{l}}$, where $H_{\mathrm{d}}$
corresponds to the isolated dots \[
H_{\mathrm{d}}=\sum_{i=1,2}(\epsilon n_{i}+Un_{i\uparrow}n_{i\downarrow})+Vn_{1}n_{2}-t\sum_{\sigma}(c_{1\sigma}^{\dagger}c_{2\sigma}+h.c.),\]
with $n_{i}=n_{i\uparrow}+n_{i\downarrow}$, $n_{i\sigma}=c_{i\sigma}^{\dagger}c_{i\sigma}$.
The dots are coupled by a tunneling matrix element $t$ and a capacitive
$V$ term. The on-site energies $\epsilon$ and the Hubbard repulsion
$U$ are taken equal for both dots. $H_{\mathrm{l}}$ describes
the noninteracting left and right tight-binding leads with hopping
parameter $t_0$ and  the coupling of the leads to the DQD. We denote
the characteristic tunneling rate of 
an isolated electron from the dot to the lead by $\Gamma=t'^2/t_0$,
where $t'$ is the parameter characterizing the dot-lead hopping. 

To calculate the ground state of the system we use the Sch\"{o}nhammer
and Gunnarsson projection-operator 
basis \cite{schonhammer76,gunnarsson85} \begin{math} |\Psi_{\lambda
  \lambda'}\rangle=P_{\lambda 1}P_{\lambda'
  2}\left|\tilde{0}\right\rangle, \end{math} 
which consists of projectors $P_{\lambda i}$; $P_{0i}=\left(1-n_{i\uparrow}\right)\left(1-n_{i\downarrow}\right)$,
$P_{1i}=\sum_{\sigma}n_{i\sigma}\left(1-n_{i\bar{\sigma}}\right)$,
$P_{2i}=n_{i\uparrow}n_{i\downarrow}$ and additional operators
involving the operators in leads. We used up to $\sim 100$ additional
combinations of operators consisting of, for example,
$P_{3i}=P_{0i}\widehat{v} P_{1i}$, where $\widehat{v}$ denotes the
tunneling to/from dot $i$. These operators are applied to the
state $\left|\tilde{0}\right\rangle$, which is the ground state
of the auxiliary noninteracting DQD Hamiltonian of the same form as
$H$, but with $U,V=0$, renormalized parameters $\epsilon, t,t'\to
\tilde{\epsilon},\tilde{t},\tilde{t}'$ and additional parameter
$\tilde{t}''$ which corresponds to hopping from left dot to
right lead and vice versa which although absent in the original
Hamiltonian is present in the effective Hamiltonian in some parameter
regimes.

The conductance is calculated using
the sine formula \cite{rejec03b},
$G=G_{0}\sin^{2}[(E_{+}-E_{-})/4t_{0}L]$, where $G_{0}=2e^{2}/h$ and
$E_{\pm}$ are the ground state energies of a large auxiliary ring
consisting of $L$ non-interacting sites and an embedded DQD, with
periodic and anti-periodic boundary conditions, respectively. 


\section{GROUND STATE AND CONDUCTANCE OF DQD WITH INTER-DOT INTERACTION}

\subsection{Detached DQDs}
The starting point towards the understanding of the ground state of DQDs are
the filling properties of isolated  DQDs ({\it
  i.e.} of the Heitler-London or the two-site Hubbard model). The first electron is
added when $\epsilon=t$, and the second when 
$\epsilon=-t+J+[(U+V)-|U-V|]/2$, where
$J=[-|U-V|+\sqrt{(U-V)^{2}+16t^{2}}]/2$ is the difference 
between singlet and triplet energies. When $n=2$ the
ground state is $[\alpha
(\ket{\uparrow\downarrow}-\ket{\downarrow\uparrow}) +
\beta(\ket{20}-\ket{02})]/\sqrt{2}$, where $\alpha/\beta =
4t/(V-U+\sqrt{(U-V)^2+16t^2})$.
The range of $\epsilon$ where single occupation is favorable is
progressively diminished when $V\neq U$. For large $t$ or at (and
near) $V=U$ the molecular bonding and anti-bonding orbitals are formed
as is seen here from $\alpha \sim \beta$.

\subsection{Attached DQDs and conductance}
\begin{figure}[h]
  \label{fig1}
  \includegraphics[width=.95\textwidth]{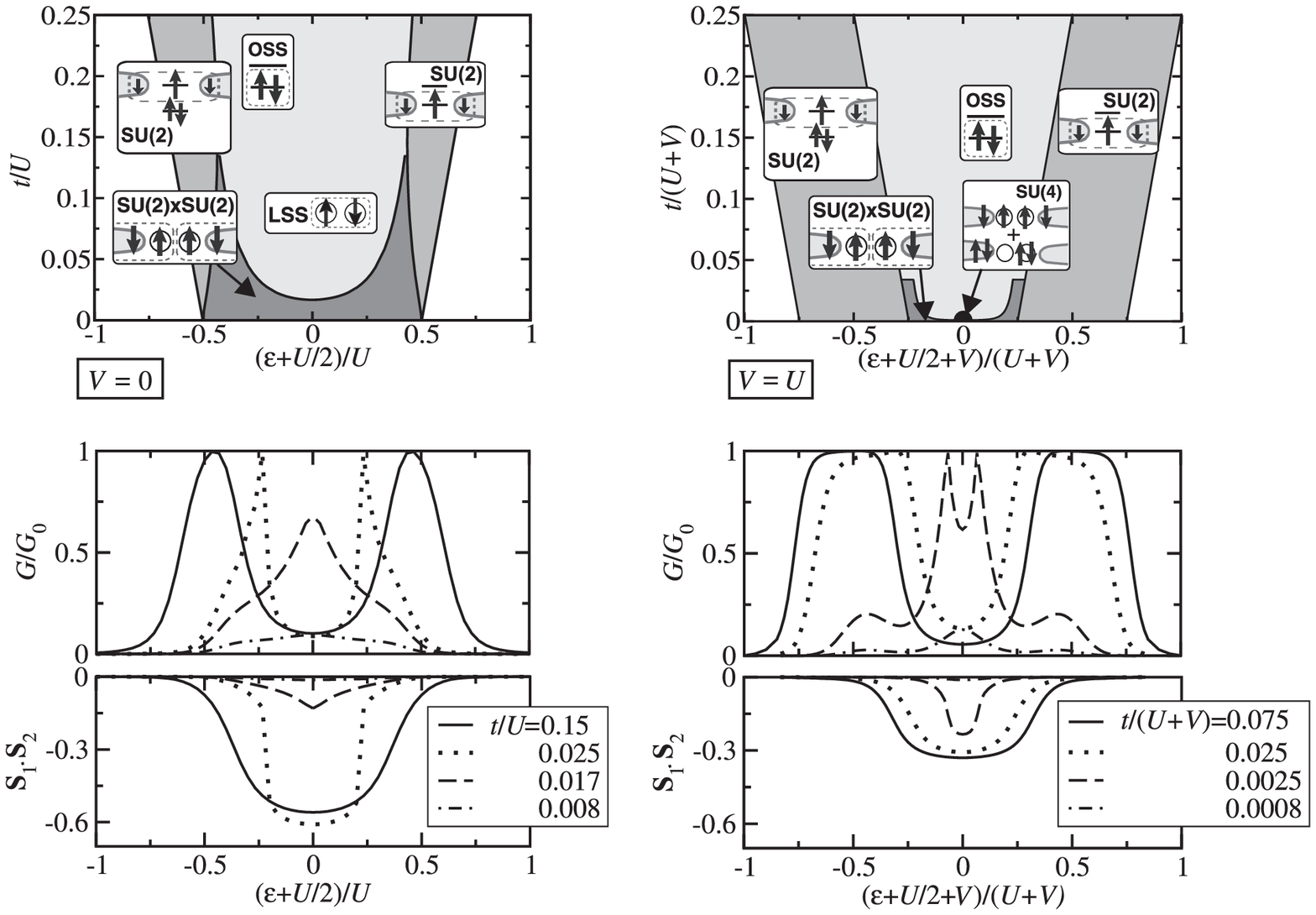}
  \caption{{\it -- Top panels}: phases of serial DQDs for $V=0$ (left) and
    $V=U$ (right). The occupancy of the DQD
    falls from left to right. Extended Kondo phases (with leads in
    pictograms) and localized
    singlet phases (without leads in pictograms) occur. {\it-- Bottom
     panels}: Conductance and spin-spin correlation of the DQD for $t$
    above (full and dotted lines for $V=0$; full, dotted and dashed
    lines for $V=U$ ) and below the localized singlet
    formation threshold. Note the approaching of
    $\mathbf{S}_1\cdot\mathbf{S}_2$ towards $-3/8$ for large $t$
  indicating the formation of the orbital singlet.}

\end{figure}

As we attach DQDs to the leads the ground state  either is or is not
reminiscent of the ground state of the isolated system. Here the latter possibility
is always due to some kind of the Kondo effect. In the top panels of
Fig.~\ref{fig1} the ground state of DQDs are
presented with pictograms for $V=0,U$ on the left and right, respectively. The near vertical dividing lines correspond to values of
parameters where the ground state of the isolated system is degenerate due
to matching energies of states with different occupancies, for example,
the rightmost line corresponds to $E(0)=0=E(1)=\epsilon-t$. The horizontal
U-shaped line is given by $J=2.2T_K$, where the scale of the Kondo
condensation energy is estimated by $T_K=\sqrt{U\Gamma/2}\exp(-\pi
\epsilon(\epsilon+U)/2\Gamma)$ for $U/\Gamma=15$. 

For $n=0,4$ interaction between electrons (or holes) is not important,
hence the ground state is not interesting. For $n=1,3$ the
ground-state of the isolated DQD is a free spin in (anti-)bonding orbital,
which is, when the leads 
are attached, at low-temperatures screened by
conduction electrons as in 'ordinary' single impurity Anderson
model. The most interesting part of the diagrams corresponds to $n\sim2 $. Here
the ground state of the isolated system is a non-degenerate singlet
but the tunneling to the leads breaks this singlet whenever roughly
twice the Kondo condensation energy exceeds the triplet excitation
energy $J$. For $V\sim U$ the $J$ is enhanced
hence the area corresponding to SU(2)$\times$SU(2) Kondo is
diminished. Near the symmetric point, however, another kind of the
Kondo effect arises for $V\sim U$ as a consequence of larger symmetry
of the $V=U$ Hamiltonian, which partially restores the occurrence of
the Kondo phase.

\subsection{Symmetries}
The Kondo effect occurs as the  consequence of the degeneracy of
states of isolated impurities. If one looks at the ground state of
two isolated impurities coupled by a capacitative (but not tunneling)
term $V=U$, one sees that the 6 states $\ket{\sigma_1 \sigma_2}$,
$\ket{20}$ and $\ket{02}$ are degenerate. Indeed, by introducing the
pseudospin operator \cite{leo04} $\tilde{T}^i=
1/2\sum_{ll'=1,2}\sum_\sigma c^{\dagger}_{l\sigma}\tau^i_{ll'}c_{l'\sigma} $, where
$\tau^i$ are the Pauli matrices, and the combined spin-pseudospin operators
$W^{ij}=S^i \tilde{T}^j$, one sees that the Hamiltonian is SU(4)
symmetric. As long as the SU(4) symmetry breaking terms are small
enough $V-U, t\lesssim T_K[SU(4)]$, the ground state is an SU(4)
'spin'  screened by the electrons in the leads.

\subsection{Orbital representation}
A complementary way is to rewrite the Hamiltonian in the basis of
orbital operators $c_{b,a}=(c_1\pm c_2)/\sqrt{2}$
\begin{equation} 
\nonumber
H_{\mathrm{d}}  =
\sum_{\alpha=a,b}\left[\epsilon_{\alpha}n_{\alpha}+\frac{U+V}{2}\left(n_{\alpha\uparrow}n_{\alpha\downarrow}+n_{\alpha\uparrow}n_{\bar{\alpha}\downarrow}\right)\right]+
V\sum_{\sigma}n_{a\sigma}n_{b\sigma}+\frac{U-V}{2}\left(C_{\textrm{flip}}-S_{\textrm{flip}}\right),
\end{equation}
where notation $\bar{a}=b$, $\bar{b}=a$ is used. The last term of
$H_{\mathrm{d}}$ consists of isospin-flip $C_{\textrm{flip}}=T_{a}^{+}T_{b}^{-}+h.c.$
and spin-flip $S_{\textrm{flip}}=S_{a}^{+}S_{b}^{-}+h.c.$ operators,
where $S_{\lambda}^{-}=c_{\lambda\downarrow}^{\dagger}c_{\lambda\uparrow}=(S_{\lambda}^{+})^{\dagger}$
are spin and $T_{\lambda}^{-}=c_{\lambda\uparrow}c_{\lambda\downarrow}=(T_{\lambda}^{+})^{\dagger}$
isospin lowering and raising operators
for the orbitals $\lambda=b,a$ (or sites $\lambda=1,2$). The full
spin (isospin) algebra is closed with operators $S_{\lambda}^{z}=(n_{\lambda\uparrow}-n_{\lambda\downarrow})/2$
and $T_{\lambda}^{z}=(n_{\lambda}-1)/2$, respectively.

\begin{figure}[t]
  \label{fig2}
  \includegraphics[width=.54\textwidth]{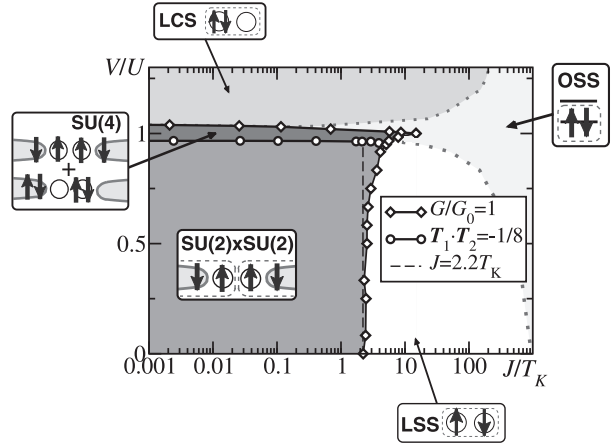}
  \caption{Phases of DQD in the point of particle-hole symmetry. The
    boundaries between Kondo and localized singlet phases are given by
    peaks in conductance and abrupt changes in correlation
    functions. The boundaries of the orbital spin singlet state are given by
    $\mathbf{S}_1\cdot\mathbf{S}_2=-3/16$ and $\Delta n^2_1=\Delta n_b^2$ on
    the upper and lower side, respectively. Note the extension of the
    Kondo phase behind the line $J=2.2T_K$ (dashed) at $V\sim U$.}
\end{figure}

When $V=U$, the spin- and isospin-flip terms in $H_{\mathrm{d}}$
are absent: the Hamiltonian is mapped exactly to the two-level Hamiltonian
with intra- and inter-level interaction $U$ with the bonding and
anti-bonding levels coupled to even and odd transmission channels,
respectively. When $V\neq U$ this mapping is no longer strictly
valid: the electrons try to avoid the inter-level repulsion by
occupying aligned spin-states in different orbitals, and the
isospin-flip terms induce the fluctuations of  charge between
orbitals. Both mechanisms prohibit electrons from occupying the well-defined orbital
states. 

\subsection{Numerical results}
In the lower panels of Fig.~\ref{fig1} the conductance and inter-dot
spin-spin correlations are plotted. Note that the orbital picture is indeed more
robust for the $V=U$ case as indicated by the broad plateaus
in conductance corresponding to the SU(2) Kondo effect of a spin residing in the
(anti-)bonding orbitals. Moreover,  $J$ is enhanced when compared to the
$V=0$ case: absence of singlet phase signalled by no peak
with unitary conductance and minor spin-spin correlation
for all $\epsilon$ occurs only for smaller $t$. Note also
that conductance is small whenever the ground state is practically
geometrically separable into  parts. In that case the flux can
be transported out of the auxiliary ring through the boundary between the parts,
yielding zero conductance in our approach \cite{rejec03b}. In
Fig.~\ref{fig2} we indicate the phases in the $(J/T_K,V/U)$
plane. Details are given in Ref. \cite{mravlje06}.

\begin{theacknowledgments}
 We acknowledge the support of SRA under grant Pl-0044.   
\end{theacknowledgments}

\bibliographystyle{aipproc} 
 \bibliography{su2su4}
\end{document}